\documentclass[aps,10pt,prc,tightenlines,twocolumn,twoside,showpacs,nofootinbib,superscriptaddress,amssymb,amsmath]{revtex4-1}
\usepackage{graphicx,amsmath,amssymb}
\usepackage{hyperref}
\usepackage{breakurl}
\usepackage{color}
\usepackage{verbatim}
\usepackage{multirow}
\usepackage{tabularx}
\usepackage{slashed}
\usepackage{soul}
\usepackage{bm} 

\newcommand{\eg}{\textit{e.g.}}
\newcommand{\etal}{\textit{et al.}}
\newcommand{\ie}{\textit{i.e.}}

\newcommand{\be}{\begin {equation}}
\newcommand{\ee}{\end{equation}}
\newcommand{\bi}{\begin{itemize}}
\newcommand{\ei}{\end{itemize}}
\newcommand{\bea}{\begin {eqnarray}}
\newcommand{\eea}{\end{eqnarray}}
\newcommand{\braket}[2]{\bra{#1}\,#2\rangle} 
\newcommand{\bra}[1]{\langle\,#1\,|}          
\newcommand{\ket}[1]{|\,#1\,\rangle}          
\newcommand{\ud}{\mathrm{d}}
\renewcommand{\b}[1]{{\bm #1}}
\newcommand{\unit}[1]{\hat {{\bm #1}}} 

\newcommand{\LCm}{{\scriptscriptstyle -}} 
\newcommand{\LCp}{{\scriptscriptstyle +}}
\newcommand{\LCpm}{{\scriptscriptstyle \pm}}

\newcommand{\LCperp}{{\scriptscriptstyle \perp}}

\begin{document}

\title{Particle distribution in intense fields in a light-front Hamiltonian approach}
\author{Guangyao Chen}   
\email{gchen@iastate.edu}
\affiliation{Department of Physics and Astronomy, Iowa State University, Ames, Iowa 50011, USA}

\author{Xingbo Zhao}
\email{xbzhao@impcas.ac.cn}
\affiliation{Institute of Modern Physics, Chinese Academy of Sciences, Lanzhou 730000, China}

\author{Yang Li}
\email{leeyoung@iastate.edu}
\affiliation{Department of Physics and Astronomy, Iowa State University, Ames, Iowa 50011, USA}


\author{Kirill Tuchin}
\email{tuchin@iastate.edu}
\affiliation{Department of Physics and Astronomy, Iowa State University, Ames, Iowa 50011, USA}

\author{James P. Vary}
\email{jvary@iastate.edu}
\affiliation{Department of Physics and Astronomy, Iowa State University, Ames, Iowa 50011, USA}

\date{\today}

\begin{abstract}
We study the real-time evolution of an electron influenced by intense electromagnetic fields using the time-dependent basis light-front quantization (tBLFQ) framework. We focus on demonstrating the non-perturbative feature of the tBLFQ approach through a realistic application of the strong coupling QED problem, in which the electromagnetic fields are generated by an ultra-relativistic nucleus. We calculate transitions of an electron influenced by such electromagnetic fields and we show agreement with light-front perturbation theory when the atomic number of the nucleus is small. We compare tBLFQ simulations with perturbative calculations for nuclei with different atomic numbers, and obtain the significant higher-order contributions for heavy nuclei. The simulated real-time evolution of the momentum distribution of an electron evolving inside the strong electromagnetic fields exhibits significant non-perturbative corrections comparing to light-front perturbation theory calculations. The formalism used in this investigation can be extended to QCD problems in heavy ion collisions and electron ion collisions.

\end{abstract}


\maketitle

\section{Introduction}
\label{sec_intro}

Real-time quantum field theory (QFT) in the non-perturbative regime is pivotal for understanding recent experimental discoveries in modern high energy nuclear colliding facilities, \eg , the Large Hadron Collider (LHC) and the Relativistic Heavy Ion Collider (RHIC). For instance, thorough understanding of jet quenching and heavy quarkonium suppression in heavy ion collisions requires detailed knowledge of how quarks and gluons interact with the evolving hot medium created by the colliding nuclei \cite{Qin:2015srf,Rapp:2008qc}. Other examples of time-dependent non-perturbative problems are, QED in ultra-intense laser fields \cite{Heinzl:2008an,DiPiazza:2011tq} and QCD in strong magnetic fields \cite{Skokov:2009qp,Tuchin:2013ie}. 

Stationary state QFT problems in the non-perturbative regime are themselves challenging, and basis light-front quantization (BLFQ) has emerged as a promising framework to solve non-perturbative QFT eigenstates from first-principles \cite{Honkanen:2010rc,Vary:2009gt,Vary:2016emi}. By employing the light-front Hamiltonian formalism, BLFQ enjoys advantages of light-front dynamics and of non-perturbative quantum many-body theory at the same time. It shares many advantageous features with discretized light-front quantization (DLCQ) \cite{Pauli:1985ps} and {\it ab initio} nuclear structure calculations, \eg, the no-core shell model (NCSM) \cite{Navratil:2000ww,Navratil:2000gs,Barrett:2013nh}. Additionally, the basis approach explicitly retains kinematical symmetries of the system and could lead to significant reduction of numerical efforts. In addition to providing the mass eigenstates, the light-front Hamiltonian formalism generates the associated light-front amplitudes which can then be applied to determine spin structures, electromagnetic form factors and generalized parton distributions of hadrons and other observables. BLFQ has been successfully applied to QED problems, \eg, the electron anomalous magnetic moment \cite{Honkanen:2010rc, Zhao:2014xaa}, the positronium system \cite{Wiecki:2014ola} and GPDs of the electron and strong coupling positronium \cite{Chakrabarti:2014cwa,Adhikari:2016idg}. Recently, BLFQ was applied to the heavy quarkonium system with a confinement potential inspired from anti-de Sitter/conformal field theory (AdS/CFT) along with the one-gluon exchange interaction from the QCD Hamiltonian \cite{Li:2015zda}. The spectroscopy and decay constants obtained from the BLFQ approach are comparable to experimental measurements and other established methods.

Distinct from the Lagrangian formulations, the Hamiltonian approach permits access to the real-time evolution of quantum states. Therefore, time-dependent basis light-front quantization (tBLFQ) is a natural extension of the BLFQ formalism. The tBLFQ formalism has been successfully applied to nonlinear Compton scattering by employing a simple {\it anzatz} for the time-dependent and intense laser field as a classical background field. Zhao \etal \, have illustrated that tBLFQ enables real-time accessibility to intermediate quantum states of the electron by showing the evolution of the invariant mass of the electron and photon Fock state, which agrees with a perturbative calculation in the small coupling limit \cite{Zhao:2013jia, Zhao:2013cma}.

It is well known that a classical description can capture the most substantial physics of the system when the occupation number is large in quantum phase space. One example is laser physics \cite{Sarachik:1970ap}. Another renowned example is the Color Glass Condensate (CGC) \cite{McLerran:1993ni,JalilianMarian:1997jx,JalilianMarian:1997gr,Iancu:2000hn}, a classical effective theory of QCD, where the small-$x$ partons are treated as classical fields generated by large-$x$ partons. Quantum effects are then treated as higher-order corrections to the classical calculations. The universal description of saturated gluons in hadrons based on CGC effective theory is able to explain a wide range of phenomena in deep-inelastic scattering and hadron-hadron collision experiments at high energies. For a recent review on this topic, see Ref.~\cite{Gelis:2010nm}. We foresee tBLFQ to be a very useful tool to study high-energy heavy ion collisions and electron ion collisions in conjunction with CGC effective theory.

As a first realistic application of the tBLFQ framework to high-energy heavy ion collisions, we investigate QED effects in which the role of strong electromagnetic (e.m.) field is yet to be understood quantitatively \cite{Tuchin:2012mf,Tuchin:2013apa,Tuchin:2013ie}. To be more specific, we study features of realistic electromagnetic fields generated by an ultra-relativistic heavy ion using the tBLFQ formalism. We solve the time-evolution of the quantized field of an electron inside such classical fields. The coupling between the electron and the strong e.m. field is at order $Z\alpha_\text{em}$ with $Z$ the atomic number of the nucleus and $\alpha_\text{em}$ the electromagnetic coupling constant. A non-perturbative approach is preferred for the strong coupling QED problem when $Z$ is large. For instance, the coupling between electron and e.m. field generated by gold nucleus is around $0.6$. We focus on demonstrating the non-perturbative features of the tBLFQ framework. This investigation also serves as a stepping-stone for future applications of tBLFQ to QCD problems in high energy nuclear collisions.

The paper is organized as follows. In section \ref{sec:bg}, we provide a brief review on the background of this investigation. Then in section \ref{sec:result} we compare the tBLFQ simulation to light-front perturbation theory (LFPT). Effects on physical observables of high energy nuclear experiments are shown in section \ref{sec:non-pert}. In section \ref{sec:conclusion}, we summarize our results and discuss additional applications of the tBLFQ framework in heavy ion collisions and electron ion collisions.

\section{Background}
\label{sec:bg}

First we briefly review the key features of BLFQ and tBLFQ, and the components of the QED Hamiltonian with classical background fields. We refer readers to Ref. \cite{Zhao:2013cma} for details. We will close this section by discussing some general properties of the intense electromagnetic fields generated by an ultra-relativistic heavy ion.

\subsection{Basis Light-front Quantization}
\label{ssec:BLFQ}

Obtaining the invariant mass eigenstates in a light-front Hamiltonian matrix approach has shown significant promise \cite{Brodsky:1997de,Wiecki:2014ola,Li:2015zda,Vary:2016emi}. The primary advantage of BLFQ is that, by adopting a basis with the same symmetries of the system under investigation, we can reduce the numerical efforts required for an accurate representation of the Hamiltonian. 

The choice of basis is arbitrary as long as it is orthogonal and complete. One of the many choices is the two dimensional harmonic oscillator (`2D-HO') basis in the transverse direction and the discretized plane-wave basis in the longitudinal direction. Each single-particle basis state can be identified using four quantum numbers, $\bar \alpha = \{k,n,m,\lambda\}$. The longitudinal momentum of the particle is characterized by the first quantum number $k$. In the longitudinal direction $x^\LCm$, we constrain the system to a box of length $2L$, and impose (anti-)~periodic boundary conditions on (fermions) bosons. As a result, the longitudinal momentum $p^\LCp=2\pi k/L$ is discretized, where the dimensionless quantity $k=1, 2, 3,...$ for bosons and $k=\frac{1}{2}, \frac{3}{2}, \frac{5}{2}, ...$ for fermions. We have neglected zero modes for bosons. The length parameter $L$ should be chosen to cover the longitudinal extent of the system, we will discuss it in section \ref{ssec:potential}. The next two quantum numbers, $n$ and $m$, depict radial excitation and angular momentum projection, respectively, of the particle within the 2D-HO basis in the transverse direction. The 2D-HO basis may be defined by two parameters, mass $M$ and frequency $\Omega$. However, we adopt a single HO parameter $b := \sqrt{M \Omega}$, since our transverse modes depend only on $b$ rather than on $M$ and $\Omega$ individually. The state carrying quantum number $n$ and $m$ has HO eigenenergy $E_{n,m}=(2n+|m|+1)\Omega$, see Appendix \ref{apd:convention} for details.

The many-particle basis states $\ket{\alpha}$ in each Fock sector are direct products of single-particle states. Such basis was initially designed for the QCD bound state problem and was supported by AdS/CFT correspondence with QCD. It has been shown, such a choice of basis allows one to encode the following three symmetries of QED Hamiltonian. First, translational symmetry in the $x^\LCm$ direction, \ie \,conservation of longitudinal momentum $P^\LCp$. Second, rotational symmetry in the transverse plane \ie \,the longituidinal projection of angular momentum $J^3 = J^3_\text{orbital} + J^3_\text{spin}$ is conserved. Finally, lepton number conservation, \ie \,net fermion number is conserved and so is the total charge. Therefore the eigenspace of QED can be grouped into segments with definite eigenvalues ${K,M_j,N_f}$ for the operators ${P^\LCp, J^3, Q}$, respectively.

The physical QED eigenstates, written as $\ket{\beta}$, are represented as the superposition of the basis states,
\begin{equation}
\ket{\beta}=\sum_{\alpha} \ket{\alpha} \braket{\alpha}{\beta}   \, ,
\label{eq:eigenbeta}
\end{equation}
with eigenstates and basis states belonging to the same segment. Coefficients $\braket{\beta}{\alpha}$ are obtained by diagonalizing $P^\LCm_\text{QED}$ in the basis representation. To this end, we require a finite dimension representation of the QED Hamiltonian that is achieved through the following basis reduction procedures.

First, by taking into account the conserved quantities and selection rules, one determines which subset of basis states will contribute to a desired observable. For this observable, one needs to work in a finite number of segments of QED eigenspace without any information loss. Second, because of the fact that even a single segment has an infinite number of degrees of freedom, we need to truncate the basis which inevitably introduces loss of precision in our calculated observables. We implement two levels of truncation scheme as follows.

{\it i) Fock-sector truncation.}
Take the physical electron state as an example. Schematically, a physical state with $N_f=1$ has the following Fock sector expansion
\begin{align}
\label{state_expan_BLFQ_2}
|e_\text{phys}\rangle=a|e\rangle+b|e\gamma\rangle+c|e\gamma\gamma\rangle+d|ee\bar{e}\rangle+\ldots \; ,
\end{align}
containing the bare electron $\ket{e}$ and its dressed states $\ket{e\gamma}$, $\ket{e\gamma\gamma}$ and $|ee\bar{e}\rangle$ etc. We explicitly assume that higher Fock-sectors give insignificant contributions to the low-lying eigenstates in which we are mostly interested, with an appropriate renormalization procedure implemented. Such assumption is motivated by the success of perturbation calculations in QFT. Furthermore, the dominance of contributions to physical observables from lower Fock sectors has been shown in scalar Yukawa model even in the non-perturbative regime \cite{Li:2015iaw}, which indirectly justifies our Fock sector truncation scheme in QED. In the following calculations we make the simplest truncation, by keeping only the single electron Fock sector in Eq.~(\ref{state_expan_BLFQ_2}), since the interactions between the fermion and the photon are suppressed by $1/Z$ comparing to the interaction between the fermion and the classical field generated by the nucleus, where $Z$ is the atomic number of the nucleus. We leave the corrections from higher Fock sectors to a future study.

{\it ii) Truncation within Fock-sectors.} Within each Fock-sector, further truncations are still needed to reduce the basis to a finite dimension. As mentioned, we impose (anti-)~periodic boundary conditions on (fermions) bosons in a longitudinal box with length $2L$. Consequently, the longitudinal momentum $p^\LCp$ of single particles can only take discrete values. We then introduce a truncation parameter $K$ on the longitudinal direction such that, $\sum_l k_l \le K$, where $k_l$ is the longitudinal momentum quantum number of $l$-th particle in the basis state. Note that systems with larger $K$ have simultaneously higher ultra-violet (UV) and lower infra-red (IR) cutoffs in the longitudinal direction. In the transverse direction, we require the total transverse quantum number $N_\alpha=\sum_l (2n_l+| m_l |+1)$ for multiparticle basis state $\ket{\alpha}$ satisfies $N_\alpha \le N_\text{max}$, where $N_\text{max}$ is a chosen truncation parameter.

We thus attain a finite dimensional representation of the QED Hamiltonian in the BLFQ basis. The continuum limit can be achieved by extending $K$ and $N_\text{max}$ to infinity. The dependence on the parameter $L$ should be weak as long as it covers the longitudinal extent of the system.

\subsection{Time-dependent Basis Light-front Quantization}
\label{ssec:tBLFQ}
The state of a quantum system at a later time is related to its state at an earlier time by the Schr\"odinger equation, which takes the following form,
\begin{equation}
  	i\frac{\partial}{\partial x^+}\ket{\psi;x^+}= \frac{1}{2}P^-(x^+)\ket{\psi;x^+} \, ,
\end{equation}
in light-front dynamics. The Schr\"odinger equation can be solved in either the interaction picture or the Schr\"odinger picture. Physical observables should not depend on the pictures we employed for the time evolution. However, for a particular problem, one picture may be numerically advantageous over another. For instance, if the Hamiltonian has a non-trivial time dependence, working in the Schr\"odinger picture may be more numerically efficient since the interaction picture would require calculating the Hamiltonian in the physical eigenstates at every time step. On the other hand, if the interaction is much smaller than the kinetic Hamiltonian, then a finer time step is required in the Schr\"odinger picture to produce the same precision as in the interaction picture. Of course the choice also depends on the physical observable of interest, \eg , the interaction picture could be more capable in describing bound states. In a word, the time evolution picture should be chosen according to the problem itself. In this investigation, since we are interested in the effects of external fields, we work in the interaction picture for time evolution. Its formal solution is,
\begin{align}
	\label{i_evolve_sch}
	\ket{\psi;x^+}_I  &= \mathcal{T}_+ \exp\bigg(-\frac{i}{2}\int\limits_0^{x^+} V_I(x^+) \bigg)\ket{\psi;0}_I \;,
\end{align}
where $\mathcal{T}_\LCp$ is light-front time ordering operator and $V_I$ is the interaction Hamiltonian in the interaction picture, with the subscript $I$ indicates the interaction picture. We can expand the initial state in the BLFQ basis,
\begin{align}
	\label{initial_c}
	\ket{\psi;0}_I = \sum\limits_\alpha \ket{\alpha} c_\alpha(0)\;,
\end{align}
where $c_\alpha(0) \equiv \braket{\alpha}{\psi;0}_I$. The coefficients of the state at later times can be expanded as
\be\label{sch-expand}
	\ket{\psi;x^+}_I  := \sum_{\alpha} c_\alpha(x^+) \ket{\alpha},
\ee
in the BLFQ basis. Its coefficients will be solved through,
\begin{align}
	\label{c_evolve}
	c(x^+) &= \mathcal{T}_+ \exp\bigg(-\frac{i}{2} \int\limits_0^{x^+} \mathcal{M}\bigg) c(0) \;.
\end{align}
where $\mathcal{M}$ is a finite dimensional representation of the Hamiltonian operator in BLFQ basis, $\mathcal{M}_{\alpha \alpha'}= \bra{\alpha} V_I \ket{\alpha'}$.  The time-evolution operator then is decomposed into small steps in light-front time $x^+$, with step size $\delta x^+$,
\be
\label{sol_m_discrete}
\mathcal{T}_\LCp \exp\bigg(-\frac{i}{2}\int\limits_0^{x^\LCp} \mathcal{M} \bigg) \rightarrow \mathcal{T}_\LCp \prod_{n} \big[1-\tfrac{i}{2}\mathcal{M}(x^\LCp_{n})\delta x^\LCp\big]  \;.
\ee
A higher-order difference scheme \cite{Askar:1978,Iitaka:1994} is implemented to ensure numerical stability and precision, refer to Appendix \ref{apd:MSD} for details. The continuum limit corresponds to the limit taking step size $\delta x^\LCp \rightarrow 0$.


\subsection{The Light-Front QED Hamiltonian}
\label{ssec:QED_Ham}

Starting from the QED Lagrangian with an additional background field,
\be
	\mathcal{L} = -\frac{1}{4}F_{\mu\nu}F^{\mu\nu} + \bar{\Psi}(i\gamma^\mu D_\mu-m_e)\Psi \;,
\ee
where $D_\mu \equiv \partial_\mu +ie C_\mu$ and $C_\mu = \mathcal{A}_\mu + A_\mu$ is the sum of the background and quantum gauge fields, respectively. In this paper, $\mathcal{A}_\mu$ is the electromagnetic field generated by the nucleus with atomic number $Z$. Note that $F_{\mu\nu}$ is calculated from $A_\mu$ alone. Working in the light-front gauge, the full Hamiltonian is then derived as \cite{Brodsky:1997de,Zhao:2013cma},
%
\begin{eqnarray}
\label{eq:FULL}
	P^- = &&\int\!\ud^2x^\LCperp\ud x^\LCm \  \frac{1}{2}\bar{{\Psi}} \gamma^\LCp \frac{m_e^2+(i\partial^\LCperp)^2}{i\partial^\LCp}\Psi  \\
	&&+ \frac{1}{2} { A}^j (i\partial^\LCperp)^2 { A}^j  + e{j}^\mu {A}_\mu  + \frac{e^2}{2} { j}^\LCp \frac{1}{(i\partial^\LCp)^2}{ j}^\LCp \nonumber \\
	&&+ \frac{e^2}{2}\, \bar{\Psi} \gamma^\mu { A}_\mu \frac{\gamma^\LCp}{i\partial^\LCp} \gamma^\nu {A}_\nu \Psi  \nonumber \\
	&&+ ej^\mu \mathcal{A}_\mu + \frac{e^2}{2}\, \bar{\Psi} \gamma^\mu \mathcal{A}_\mu \frac{\gamma^\LCp}{i\partial^\LCp} \gamma^\nu \mathcal{A}_\nu \Psi \nonumber \\
	&&+\frac{e^2}{2}\, \bar{\Psi} \gamma^\mu { A}_\mu \frac{\gamma^\LCp}{i\partial^\LCp} \gamma^\nu \mathcal{A}_\nu \Psi + \frac{e^2}{2}\, \bar{\Psi} \gamma^\mu \mathcal{A}_\mu \frac{\gamma^\LCp}{i\partial^\LCp} \gamma^\nu {A}_\nu \Psi \;.  \nonumber
\end{eqnarray}

%
The first three lines are the QED light-front Hamiltonian, $P^-_\text{QED}$. In order, each of the first five terms in Eq.~(\ref{eq:FULL}) represents the fermion kinetic energy $T_f$, photon kinetic energy $T_\gamma$, {\it vertex interaction} $W_1$, {\it instantaneous-photon interaction} $W_2$ and {\it instantaneous-fermion interaction} $W_3$ respectively. The last two lines contain the four new interactions generated by the classical background field $\mathcal{A}$, we label them as $\mathcal{W}_1$, $\mathcal{W}_2$, $\mathcal{W}_3$ and $\mathcal{W}_4$ respectively. Since we only keep the leading Fock-sector which contains one single fermion, only $T_f$, $\mathcal{W}_1$ and $\mathcal{W}_2$ enter our calculation. In the following, we only keep the relevant terms in the QED Hamiltonian,
 \begin{equation}
   P^-=P^-_\text{QED} + \mathcal{V} (x^\LCp) \, ,
   \label{eq:H_TandV}
   \end{equation}
where $P^-_\text{QED}=T_f$ and $\mathcal{V} (x^\LCp)=\mathcal{W}_1 (x^\LCp) +\mathcal{W}_2(x^\LCp)$ throughout our discussion here. Note $\mathcal{W}_1$ is first order in $Z \alpha_\text{em}$ and $\mathcal{W}_2$ is second order in $Z \alpha_\text{em}$, with $\alpha_\text{em} \sim 1/137$ being the electromagnetic coupling constant.

Our particular truncation of Fock-sector also simplifies the problem as we can take physical values for the electron mass and charge. If one works with higher Fock-sectors with both electron and photon, proper renormalization is required. One feasible choice is the sector dependent scheme \cite{Karmanov:2008br,Karmanov:2012aj,Chabysheva:2009ez}, which has been applied to the QED Hamiltonian when calculating the electron anomalous magnetic moment, for which the result agrees with the Schwinger value within 1\%~\cite{Zhao:2014xaa}.

\subsection{Electromagnetic Fields Generated by Relativistic Heavy Ion}
\label{ssec:potential}
The charge densities and current densities of one ion with atomic number $Z$ moving along the $z$-axis with velocity $v$ are
\begin{eqnarray}
\rho(z,t,\b x^\perp)=Z \vert e \vert \delta(z-vt)\delta(\b x^\perp) \;,  \nonumber \\
 {\b j} (z,t,\b x^\perp)=Z \vert e \vert v \unit z \delta(z-vt)\delta(\b x^\perp) \; .
\end{eqnarray}
The four vector potential of the fields obeys,
\begin{eqnarray}
(\nabla^2 - \partial_t^2) \mathcal{A}^0 = -\rho   \, , \nonumber \\
(\nabla^2 - \partial_t^2)  \mathcal{\b A} = - \b j   \, ,
\end{eqnarray}
where we omit the vacuum permittivity and permeability in natural units. In the light-cone gauge, in terms of $k^\LCp,\b k^\perp,x^\LCp$, the solutions of the above equations are,
\begin{eqnarray}
\mathcal{A}^- =&     2  Z e  \frac{e^{-2y} }{ \left(  e^{-2y} (k^\LCp)^2 +  \bm k_\perp^2 \right)} e^{\frac{i}{2} e^{ -2y } k^+ x^+ }    \, , \nonumber \\
\mathcal{A}^i =& -  Z e \frac{k^i}{k^+}  \frac{ 1 }{ \left( e^{-2y} (k^\LCp)^2 + \bm k_\perp^2 \right)} e^{\frac{i}{2} e^{ -2y } k^+ x^+}    \, ,
\label{eq:4vector}
\end{eqnarray}
where $y=\frac{1}{2} \ln(P^\LCp/P^\LCm)$ is the rapidity of the heavy ion, with $P^\mu$ the momentum four-vector of the heavy ion.

\begin{figure}[!t]
\centering
\includegraphics[width=0.47\textwidth]{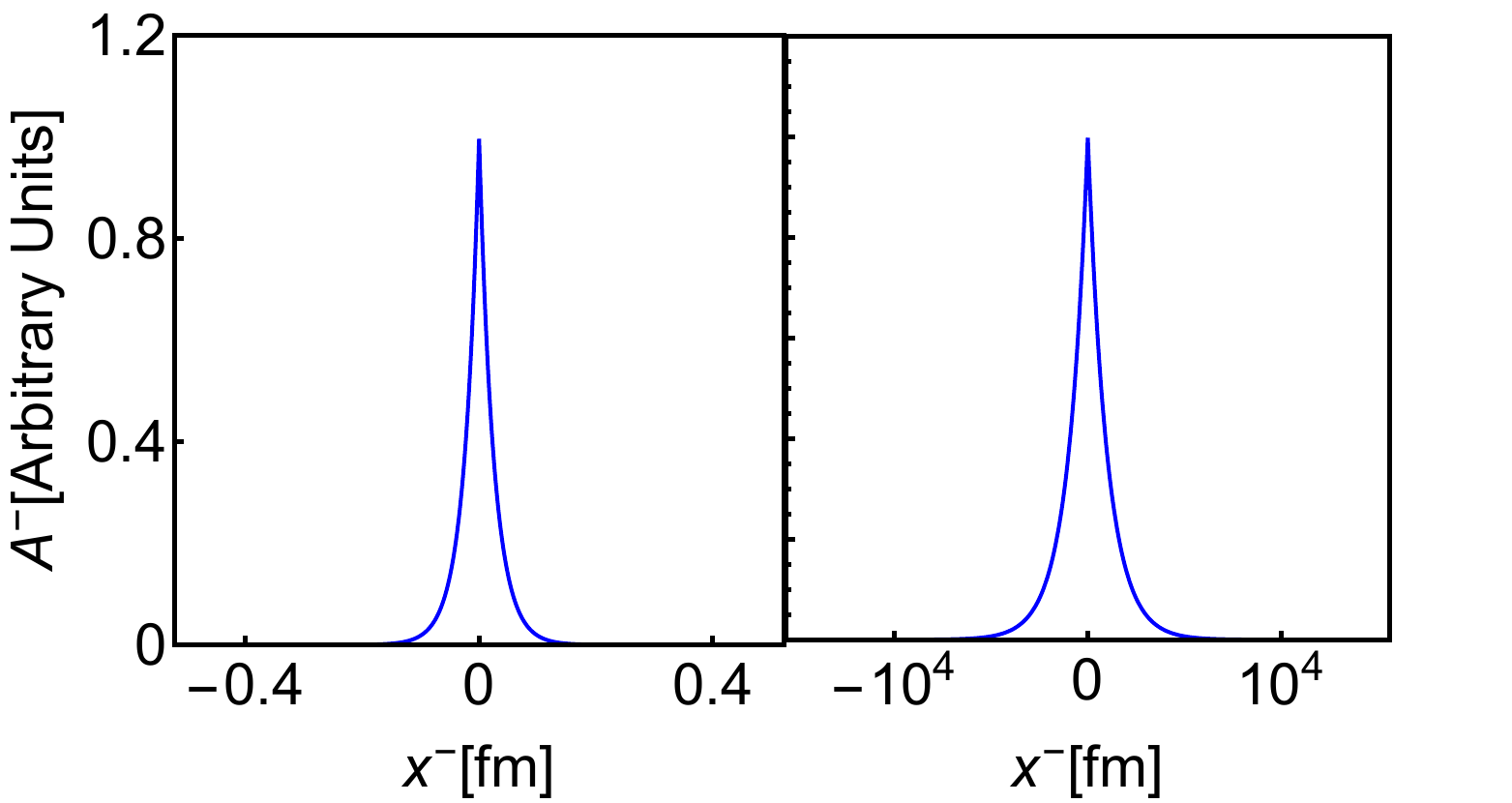}
\caption{A typical $x^\LCm$ distribution of the $-$ component of the potential, $\mathcal{A}^\LCm$, produced by a heavy ion moving along the positive (left) and the negative (right) $z$-axis in rapidity $y=\pm 5.3$ at $x^\LCp=0$, for the modes with transverse momentum $k_\perp=1$~GeV. The potential as a function of $x^\LCm$ can be obtained through a Fourier transformation of Eqs.~(\ref{eq:4vector}).
The width of such distribution increases as the transverse momentum decreases.
}
\label{pic:Apos}
\end{figure}

Let us discuss the spatial distribution of the potential before we proceed with our calculation. In modern high energy collision facilities, particles are accelerated to the ultra-relativistic regime. For example, at RHIC, the center of mass energy of the collisions reaches $200$~GeV per nucleon, the rapidity of the colliding nuclei is $y \approx 5.3$. The energy is even higher at the LHC, \eg \, the rapidity of the colliding particles are $y \sim 9.5$ at center of mass energy $1.4$~TeV. Thus $e^{-2y}$ is a small (large) number when the heavy ion is moving along positive (negative) $z$-axis. Apparently, the potential generated by a heavy ion moving along the positive $z$-axis is almost stationary with a period $2\pi e^{2 \vert y \vert }/k^\LCp$ in $x^\LCp$, and it has a very narrow extent in the longitudinal direction, see Fig. \ref{pic:Apos} . Contrarily, the potential generated by a heavy ion moving along negative $z$-axis is oscillating rapidly with a period $2\pi e^{-2 \vert y \vert }/k^\LCp$ in $x^\LCp$, while it has a very wide extent in the longitudinal direction, see Fig. \ref{pic:Apos}.

Although the potential profiles for a heavy ion moving along the positive and negative direction are very different from a first look, physical observables for the same process must be independent of such mathematical treatment. In tBLFQ, to achieve an accurate description of the same process, a larger coverage in $x^\LCp$ ($x^\LCm$) is necessary for a heavy ion moving in the positive (negative) $z$-direction and their continuum limits should be equivalent. In the following discussion, we assume the heavy ion is moving along the positive $z$-axis. First, it is easier to handle the time evolution of a \textit{quasi stationary} potential. Second, the potential is concentrated in a smaller region in $x^\LCm$. Consequently a moderate truncation parameter $L$ is sufficient to enclose the potential and cover a wide region in longitudinal momentum at the same time.

We are now ready to discuss the truncation parameter $L$ introduced in section \ref{ssec:BLFQ}. The guidelines are that, the potential outside the box only makes inconsequential corrections to the process we are interested in, while at the same time a smaller $L$ provides larger longitudinal momentum coverage for the same truncation parameter $K$. Eqs.~(\ref{eq:4vector}) clearly suggest that the longitudinal extent of the potential in $x^\LCm$ depends on the transverse momentum we are interested in. We will specify our choice of $L$ for each calculation we perform in the following.

\section{Comparison to Light-Front Perturbation Theory}
\label{sec:result}

The equivalence of the LFPT and the covariant perturbation theory has been established decades ago \cite{Chang:1973qi}. Furthermore, QED in the perturbative regime has been verified up to very high precision by various experiments \cite{Mohr:2008fa}. Thus we can check the validity of our formalism using LFPT in the perturbative regime and study the numerical error introduced by the truncations and time-step discretization. In addition, we can study higher-order contributions by comparing tBLFQ simulations to LFPT calculations.

\subsection{Comparison to Momentum Basis}
\label{ssec:ham_sch}

The Hamiltonian matrix elements $\bra{\alpha'} \mathcal{V} \ket{\alpha}$ for the potential in Eq.~(\ref{eq:4vector}) in the BLFQ basis can be calculated algebraically and the detailed expressions are presented in Appendix \ref{apd:Ham}. The first check would be that the transition amplitudes induced by the interaction $\mathcal{V}$ between particular initial and final states are consistent in the BLFQ basis and momentum basis, with sufficiently large $N_\text{max}$. We compare the leading order in the coupling between electron and background field $\alpha \equiv Z \alpha_\text{em}$ so that only $\mathcal{W}_1$ is relevant. In the longitudinal direction we adopt the discretized momentum basis, with which we approach the continuum longitudinal momentum limit when $K$ increases. For simplicity in the HO basis, we adopt a wave packet which is a Gaussian in the transverse direction. The Gaussian packet is centered at $\b p_0^\perp$ and the width of the Gaussian wave packet $\sigma_0$ can be chosen independent of the 2D-HO parameter in the BLFQ basis. When $\sigma_0 \rightarrow 0$ the initial state becomes a representation of the transverse momentum eigenstate. In the longitudinal direction we use its momentum eigenstate. Thus, the initial state is labeled by the following quantum numbers,
\begin{equation}
 \ket{\phi_0} = \ket{p^\LCp_0,G(\sigma_0,\b p_0^\perp),\lambda_0} \; .
\end{equation}
Its normalized wavefunction in the transverse direction is,
\begin{equation}
\phi_0^\perp (\b p^\perp)=\braket{\b p^\perp}{G(\sigma_0,\b p_0^\perp)} = \frac{ 1 }{\sqrt{\pi} \sigma_0} e^{-\frac{(\b p^\perp-\b p^\perp_0)^2}{2\sigma_0^2}}  \; .
\end{equation}
The transition amplitude $\bra{\phi_f} \mathcal{W}_1 \ket{\phi_i}$ to a momentum eigenstate,
\begin{equation}
 \ket{\phi_f} = \ket{p^\LCp_f,\b p_f^\perp,\lambda_f} \; ,
\end{equation}
can be calculated by integrating over the initial transverse momentum distribution,
\begin{eqnarray}
 && \int \ud^2 \b p^\perp \bra{p^\LCp_f,\b p_f^\perp,\lambda_f} \mathcal{W}_1 \ket{p^\LCp_0,\b p^\perp,\lambda_0} \phi_0^\perp (\b p^\perp)  
 \; .
\label{eq:ham_mom}
\end{eqnarray}
It can also be calculated in the BLFQ basis as follows,
\begin{eqnarray}
\sum_{\alpha,\alpha'} \braket{p^\LCp_f,\b p_f^\perp,\lambda_f}{\alpha'} \bra{\alpha'} \mathcal{W}_1 \ket{\alpha} \braket{\alpha}{p^\LCp_0,G(\sigma_0,\b p_0^\perp),\lambda_0} \; , \nonumber \\
\label{eq:ham_ho}
\end{eqnarray}
where $\braket{\alpha}{p^\LCp_0,G(\sigma_0,\b p_0^\perp),\lambda_0}$ can be calculated analytically. In principle, Eqs.~(\ref{eq:ham_mom}) and (\ref{eq:ham_ho}) are identical if we sum over all $\alpha$ states. In practice, we can only perform calculations using the truncation scheme explained in section \ref{ssec:BLFQ}. It is then necessary to check the behavior of the numerical uncertainty as we increase our truncation parameters.

\begin{figure*}
\centering
\includegraphics[width=0.47\textwidth]{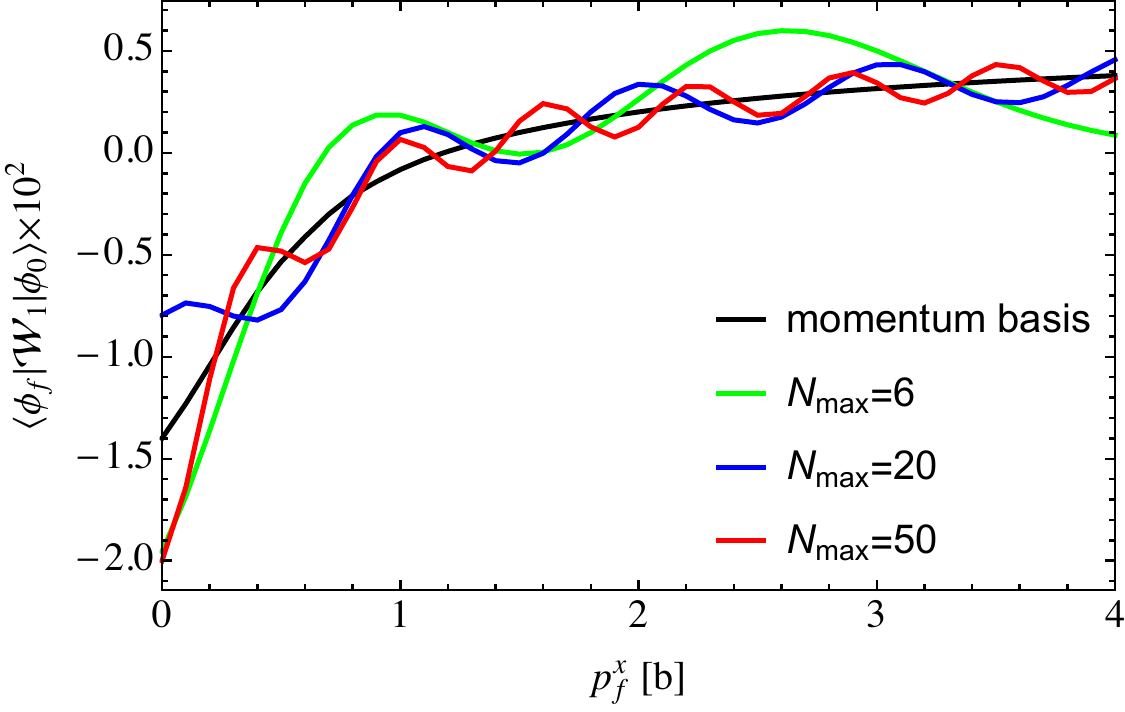}
\includegraphics[width=0.47\textwidth]{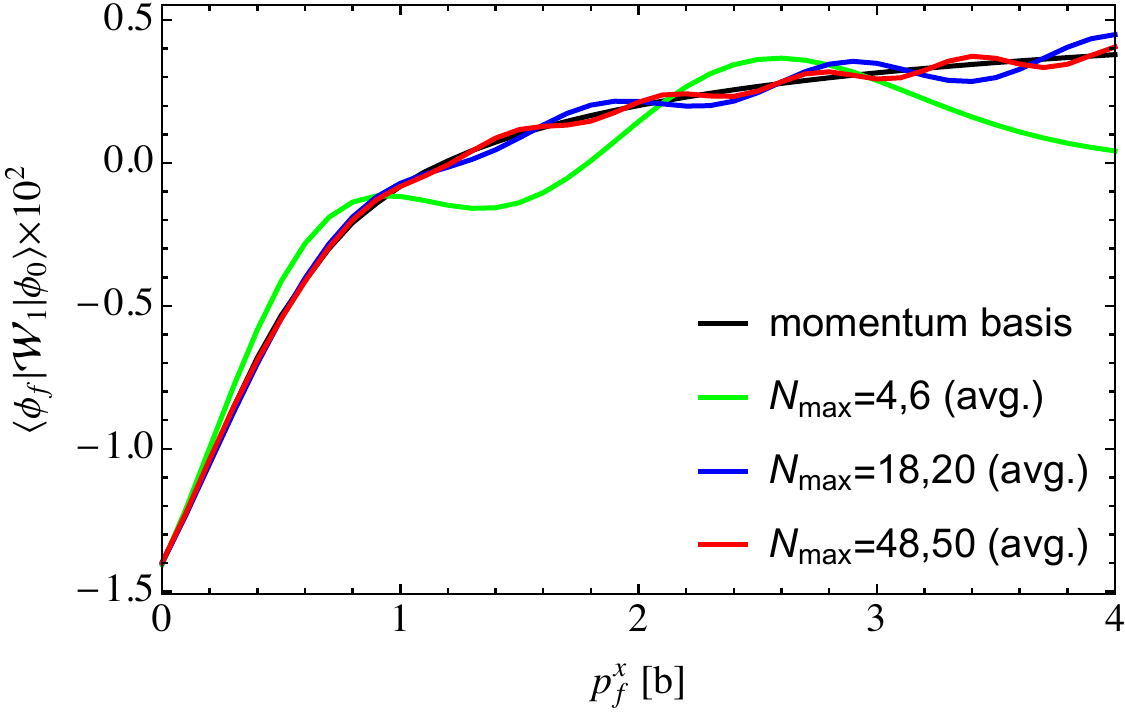}
\caption{Comparison of transition amplitude $\bra{\phi_f} \mathcal{W}_1 \ket{\phi_0}$ in momentum basis and BLFQ basis as a function of $p_f^x$ in units of the HO length scale $b$. Initial state $\phi_0$ has longitudinal momentum $p^\LCp_0=3\pi/10$~GeV, and helicity $\lambda_0=1/2$; in transverse direction, it is a Gaussian wave packet centered at $\b p_0^\perp = (b/2,0)$ with width $\sigma_0=b/2$. Final state $\phi_f$ has longitudinal momentum $p^\LCp_f=7\pi/10$~GeV, $\lambda_f=1/2$ and $p_f^y=0$. HO parameter $b=1000 m_e$. \textit{Left}: Comparison with momentum basis transition amplitude using BLFQ basis calculation at different $N_\text{max}$. \textit{Right}: Comparison with momentum basis transition amplitude using BLFQ basis calculation averaged over two $N_\text{max}$. 
}
\label{pic:rawfit}
\end{figure*}

We study the transition amplitude of an electron influenced by the field generated by a gold nucleus moving along the positive $z$-axis with rapidity $y=5.3$. We study transition amplitudes calculated in the BLFQ basis and the momentum basis from various initial states and final states, the comparisons show similar convergence behavior. As an example, we consider the initial state $\ket{\phi_0}$ at $x^\LCp=0$ with $p^\LCp_0=3\pi/10$~GeV, $\lambda_0=1/2$, and a Gaussian wave packet centered at $\b p_0^\perp = (b/2,0)$ with width $\sigma_0=b/2$. We set the HO parameter in the BLFQ basis to $b=1000 m_e = 0.511$~GeV. We calculate the amplitude to the final state $\ket{\phi_f}$ which has the following quantum number, $p^\LCp_f=7\pi/10$~GeV,$\lambda_f=1/2,\b p_f^\perp = (p^x_f,0)$ in the BLFQ basis with truncation $N_\text{max}$ and compare it to a momentum basis calculation.  Fig. \ref{pic:rawfit} shows the comparison between the BLFQ basis results with different $N_\text{max}$ and the momentum basis calculation. The difference between those two bases decreases as we increase $N_\text{max}$. However, even for $N_\text{max}=50$, we find their discrepancy is not negligible. On the other hand, if we average results of two BLFQ basis calculations with consecutive even $N_\text{max}$, the results show excellent agreement with the momentum basis calculation at $N_\text{max}=50$. This indicates averaging over $N_\text{max}$ can effectively enhance the convergence to momentum basis calculation.

Mathematically, the challenge in comparing HO basis and momentum basis calculation is rooted in the fact that HO functions are square-integrable while plane waves are not. Intuitively, the averaging procedure we adopted above can be illuminated by the following example. It is well known that a 2D delta function has its 2D-HO wavefunction representation,
\begin{equation}
\sum_{n,m}  \tilde\Phi^*_{n m}(\b p^\perp) \tilde\Phi_{n m}(\b q^\perp) = (2\pi)^2\delta^{(2)} (\b p^\perp-\b q^\perp) \; ,
\end{equation}
the equality is exact if $n$ and $m$ are summed over all possible values. However if we constrain the domain of $n$ and $m$ by requiring $N_\text{total}=2n+\vert m \vert+1 \le N_\text{max}$, the following integral
\begin{equation}
\int \frac{ \ud^2 \b p^\perp}{(2\pi)^2} \sum_{n,m} \tilde\Phi^*_{n m}(\b p^\perp) \tilde\Phi_{n m}(\b 0)
\end{equation}
is oscillating between the discrete results $0$ and $2$ as $N_\text{max}$ increases. Such a behavior provides an argument for averaging over consecutive $N_\text{max}$ results when comparing HO basis calculations to momentum basis calculations.

We have shown that when $N_\text{max}$ and $K$ are sufficiently large, the BLFQ basis results exhibit good agreement with the momentum basis results. At this stage, we can now concentrate on the behavior of discretized numerical time evolution scheme where all calculations are performed in the BLFQ basis.

\subsection{Transition Rate}
\label{ssec:rate}

\begin{figure*}[t!]
\centering
\includegraphics[width=0.9\textwidth]{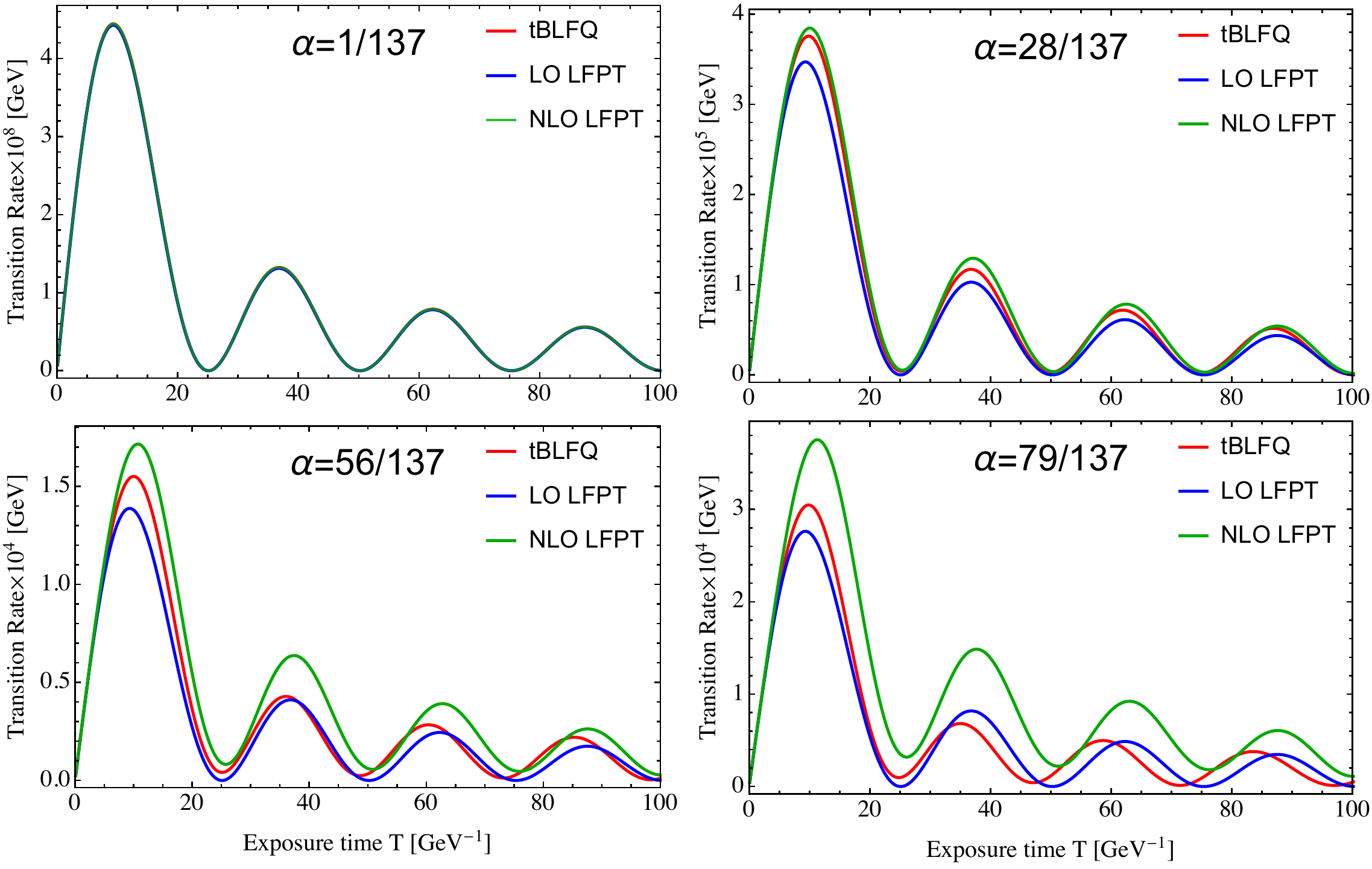}
\caption{\label{fig:rate}  (Color online) Transition rate as a function of exposure time $T$ for  an electron between a specific initial and a specific final state induced by the potential which is generated by nuclei with atomic number $Z=1,28,56,79$, the coupling between electron and background field $\alpha \equiv Z \alpha_\text{em}$ with $\alpha_em = 1/137$. The nucleus is moving along the positive $z$-axis with $y=5.3$. The initial and final states are kinetic energy eigenstates with energies $P^\LCm_{\beta,i}=0.455$~GeV and $P^\LCm_{\beta,f}=0.955$~GeV. The calculation is performed using $N_\text{max}=32$, $K=32$, $L=10$~GeV$^{-1}$ and $b=1000 m_e$, see text for details.
}
\end{figure*}

Perturbation theory provides contributions from the interaction through a power series in the coupling constant $\alpha$, which is expected to converge for sufficiently small $\alpha$. Applying perturbation theory to the S-matrix, we have,
\begin{equation}
\mathcal{S}=1-\frac{i}{2} \lim_{T \rightarrow \infty} \int_{0}^{T} d x^\LCp P^\LCm_{int} (x^\LCp)+ ...
\end{equation}
where we have only kept the leading order (LO) contribution in $\alpha$. Contrary to the infinite time limit in perturbation theory, we can only evolve for a finite time using the discretized numerical time evolution scheme described in Eq.~(\ref{c_evolve}). Extrapolation to the infinite time limit requires infinite energy resolution which can be achieved only in the limit with $N_\text{max}$ and $K$ approaching infinity. With finite $N_\text{max}$ and $K$ we make a compromise by comparing to the perturbative calculation without taking the infinite time limit.

The most important physical observable measured in scattering experiments is the cross section, which is related to the transition rate by a flux factor. The transition rate is defined as the transition probability from state $\ket{i}$ to $\ket{f}$ divided by the time $T$ during which the interaction is active,
\begin{equation}
\Gamma_T(i \rightarrow f)=\frac{P_T(i \rightarrow f)}{T} \; ,
\end{equation}
where $P_T(i \rightarrow f) = \vert \bra{f} \mathcal{S} \ket{i} \vert^2 $.
The scattering matrix can be calculated using either LFPT or the tBLFQ formalism. One advantage of the tBLFQ approach is that it does not rely on any expansion in the coupling constant of the interaction thus it is a legitimate approach for both weak and strong interactions. The applicability of tBLFQ for time-dependent non-perturbative problems is rooted in the numerical schemes we adopted for basis construction and time evolution. We have discussed extensively the non-perturbative feature of the BLFQ approach in section \ref{ssec:BLFQ}. Therefore, we now focus on how non-perturbative effects are incorporated by our discretized time evolution scheme.

Conceptually, all numerical time evolution schemes are implemented by decomposing the time evolution operator into many small steps with step size ($\delta x^\LCp$) in light-front time $x^\LCp$,
\begin{equation}
\label{sol_S_discrete}
\mathcal{T}_\LCp e^{-\frac{i}{2}\int\limits_0^{x^+} P^\LCm}\rightarrow \mathcal{T}_\LCp \prod_{n} \big[1-\tfrac{i}{2} P^\LCm (x^+_{n})\delta x^+\big] \; . 
\end{equation}
Contributions from up to $n$-th order in $\alpha$ are preserved, with $n$ the total number of time steps. The resummation up to all orders in $\alpha$ is achieved by taking the limit $\delta x^\LCp \rightarrow 0$.

As a demonstration of the non-perturbative feature of the tBLFQ approach, we consider the transition of an electron between two physical QED eigenstates $\ket \beta$, as defined in Eq.~(\ref{eq:eigenbeta}), in the fields generated by different nuclei. We construct the BLFQ basis states as follows. We require the electron to be in the segments with definite eigenvalues $m=0$ and $\lambda=1/2$. We can make such a choice for the following two reasons. First, the potential is azimuthally symmetric, thus the total angular momentum projection $M_j = m+ \lambda$ of the evolving electron is conserved. Second, the spin projection of the electron $\lambda$ is approximately conserved, since the helicity flip processes are suppressed by $m_f/p^\perp$ comparing to the helicity non-flip processes, where $m_f$ is the mass of the fermion and $p^\perp$ is its typical transverse momentum, see Table.~\ref{tab:spinor_JA_contraction_list} for detail. We set the longitudinal box length to be $L=10$~GeV$^{-1} \approx 2$~fm. As discussed above, this is sufficient for the potential in our application. We set the HO parameter to be $b=1000 m_e=0.511$~GeV, which is chosen as a representative of the typical transverse momentum of particles observed in heavy ion collisions. The BLFQ basis is constructed using $N_\text{max}=32$, $K=32$. The physical QED eigenstates are then obtained by diagonalizing the fermion kinetic energy in this BLFQ basis.

We take each nucleus to be moving along the positive $z$-axis with $y=5.3$. Since the nucleus is moving almost along $x^\LCp$, we approximate the generated potential as a static potential during the time interval our calculation is performed. We then assume an electron enters such fields at time $x^\LCp=0$ in a physical QED $\ket{\beta_i}$ of $P^\LCm_\text{QED}$, with energy $P^\LCm_{\beta,i}=0.455$~GeV, which belongs to the segment of BLFQ basis states with $k=43/2$. We calculate the transition rate of the electron to the final state $\ket{\beta_f}$ with energy $P^\LCm_{\beta,f}=0.955$~GeV, which belongs to the segment of BLFQ basis states with $k=45/2$. We have only considered the transition induced by the $\mathcal{W}_1$ term of the Hamiltonian. In Fig.~\ref{fig:rate}, we show the transition rate from the initial eigenstate $\ket{\beta_i}$ with energy $P^\LCm_{\beta,i}=0.455$~GeV to the final state $\ket{\beta_f}$ with energy $P^\LCm_{\beta,f}=0.955$~GeV, for nuclei with atomic number $Z=1,28,56,79$ as a function of exposure time T.

We make the following observations. First, in the small coupling regime, the tBLFQ calculation agrees with the NLO LFPT calculation, and both of them are only slightly different from the LO LFPT calculation, see the case with $\alpha=1/137$ in Fig. \ref{fig:rate}. Such an agreement confirms the equivalence of the tBLFQ approach and the LFPT calculation in the small coupling regime. Second, in the strong coupling regime, both the tBLFQ and the NLO LFPT calculations dramatically differ from the LO LFPT calculation, see the case with $\alpha=79/137$ in Fig.~\ref{fig:rate}. Note that the period of the transition rate as a function of exposure time has changed due to higher order effects. The tBLFQ calculation should be regarded as a good approximation of the all order resummation results. Thus our comparison at strong coupling indicates that higher-order effects are significant for the interaction between the charged fermion and the electromagnetic field generated by an ultra-relativistic heavy ion. The plots in Fig.~\ref{fig:rate} also display the anticipated growth of higher-order effects with increasing atomic number of the nucleus.

\section{Physical Observables}
\label{sec:non-pert}
%
The electromagnetic field strength immediately after an ultra-relativistic heavy ion collision is proportional to the collision energy and reaches $m_{\pi}^2$ at RHIC and $10 m_{\pi}^2$ at LHC. In addition, the field in the QGP medium could last up to a few fm/c. Such strong fields could lead to major modifications of physical observables, \ie \, the flow of the QGP, particle production, heavy quarkonium dissociation and so on, see Ref.~\cite{Tuchin:2013ie} for a review. In principle such modifications are within reach using the tBLFQ framework. However, as discussed in Sec. \ref{ssec:potential}, the fields generated by two colliding heavy ions moving in opposite directions have different dependence on light-front coordinates, thus it is a numerical challenge to study them simultaneously at this stage. In this paper, we study the real-time evolution of the momentum distribution of a fermion evolving inside the strong electromagnetic fields generated by one relativistic heavy ion as a demonstration of the tBLFQ formalism.


\begin{figure}[t!]
\centering
\includegraphics[width=0.40\textwidth]{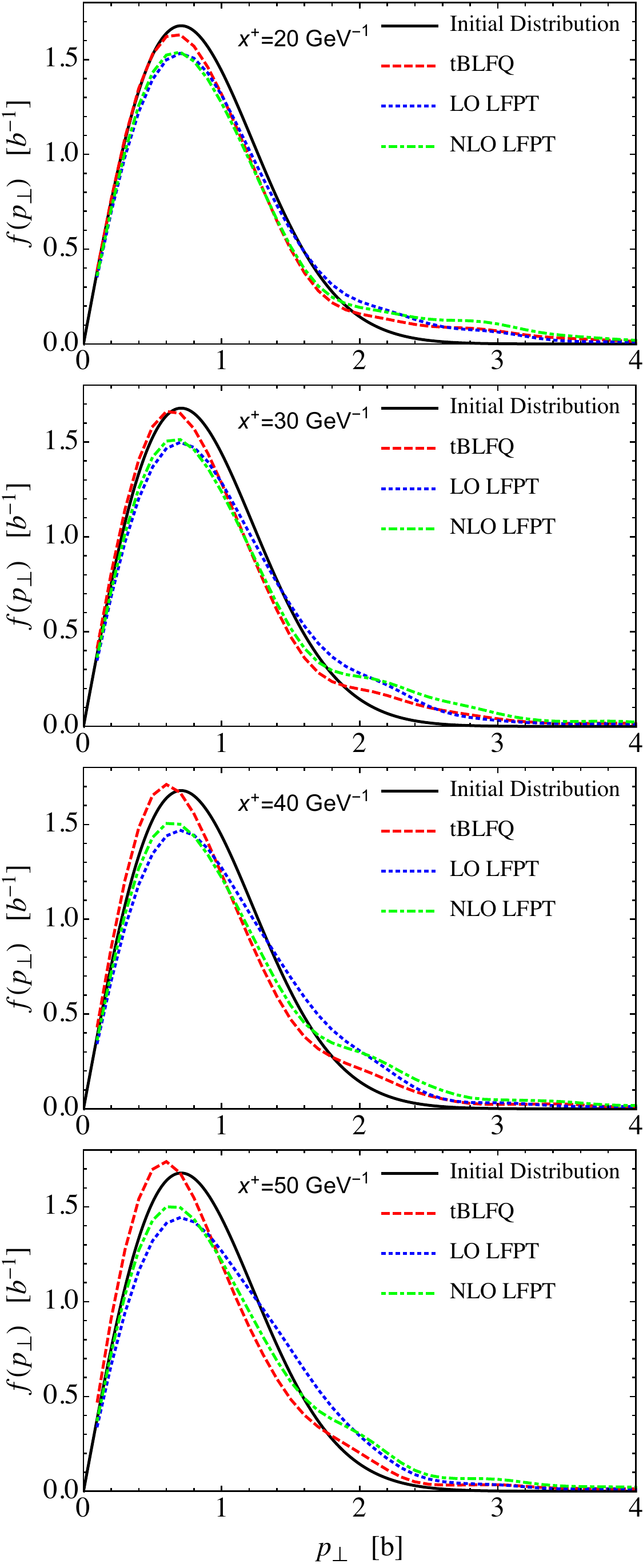}
\caption{\label{fig:nonpert2}  Snapshots in time $x^\LCp$ of the transverse momentum distribution of an electron inside the e.m. field generated by a gold nucleus moving along the positive $z$-axis with rapidity $y=5.3$. The initial state of the electron is the BLFQ basis state with $k=\frac{17}{2}$, $n=0$, $m=0$ and $\lambda=\frac{1}{2}$ at $x^\LCp=0$. Note that the chosen initial state is not an eigenstate of the pure BLFQ Hamiltonian. The calculation is performed using $N_\text{max}=32$, $K=32$, $L=10$~GeV$^{-1}$ and $b=1000 m_e$. 
}
\end{figure}

\begin{figure}[t!]
\centering
\includegraphics[width=0.45\textwidth]{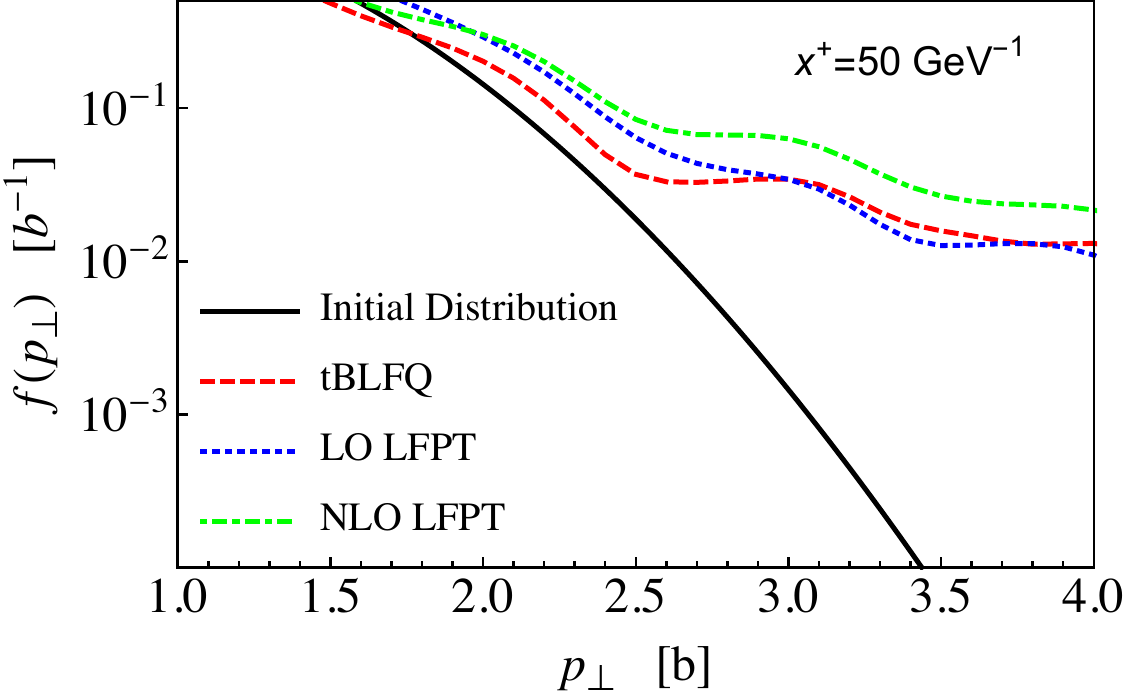}
\caption{\label{fig:nonpert4}  Snapshot of transverse momentum distribution on a semi-log scale for the same process in Fig.~\ref{fig:nonpert2} at $x^\LCp=50$~GeV$^{-1}$.}
\end{figure}

The momentum distributions of produced particles carry rich information about the collision process. In heavy ion collisions, the momentum distributions of various probes provide tomographic properties of the hot and dense medium created. Comparing to hadronic observables, electromagnetic probes, such as direct photons and dileptons, are valued for their greatly reduced final-state interactions. However, the strong magnetic field generated by the relativistic heavy ion \cite{Skokov:2009qp} could significantly modify the configurations of charged particles, especially if they are produced immediately after the collision when the magnetic field is still strong \cite{Tuchin:2013ie}. In this section we demonstrate real-time evolution of the momentum distribution for an electron evolving inside the strong electromagnetic field generated by an ultra-relativistic heavy nucleus using the tBLFQ formalism.

Consider the case in which a gold nucleus is moving along the positive $z$-axis with rapidity $y=5.3$. We perform our calculation starting at time $x^\LCp=0$, at which the electron is in the BLFQ basis state $\ket{\alpha_0}$ with the following quantum number, $k=\frac{17}{2},n=0,m=0,\lambda=\frac{1}{2}$. As a result the initial longitudinal momentum of the electron is $p^\LCp=17\pi/L$, and the initial transverse momentum distribution of the electron is,
\begin{equation}
   f_0 (p^\perp) = \frac{p^\perp}{2\pi}  \vert \tilde \Phi^b_{00} (\bm p^\perp)\vert^2 \; ,
\end{equation}
where $p^\perp \equiv \vert \b p^\perp \vert$ and the transverse distribution is normalized as $\int \ud p^\perp  f_0 (p^\perp) =1$. The longitudinal length was chosen to be $L=10$~GeV$^{-1}$, such that the longitudinal extent of the potential is limited. The HO parameter is set to be $b=1000 m_e=0.511$~GeV in light of the fact that, in heavy ion collisions, the transverse momentum of the e.m. probes could reach a few GeV.

\begin{figure}[t!]
\centering
\includegraphics[width=0.40\textwidth]{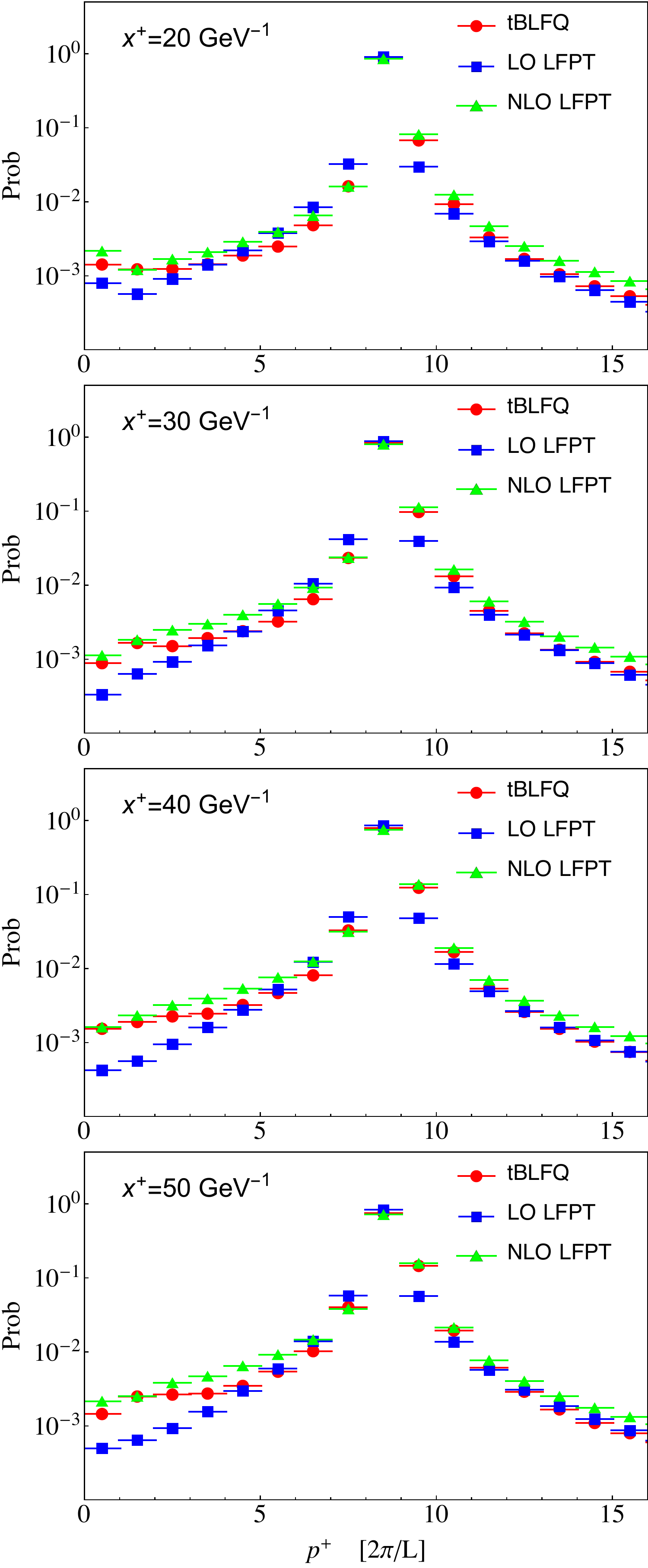}
\caption{\label{fig:nonpert3}  Snapshots of longitudinal momentum distribution for the same process in Fig.~\ref{fig:nonpert2}. Horizontal bars indicate the momentum bin width for the discretized plane waves in the longitudinal direction.}
\end{figure}

Using the tBLFQ formalism, we have access to the real-time evolution of the configuration of particles at the amplitude level. In this section, we illustrate this by showing the momentum probability distribution of the electron in both transverse and longitudinal directions as a function of time. We have considered the transition induced by both $\mathcal{W}_1$ and $\mathcal{W}_2$ terms of the Hamiltonian. We show snapshots of the transverse momentum probability distribution integrated over longitudinal momentum in Fig.~\ref{fig:nonpert2} and the longitudinal momentum probability distribution in Fig.~\ref{fig:nonpert3}, simulated in tBLFQ and compared to the LO and NLO LFPT calculations using truncation parameters $N_\text{max}=32$ and $K=32$. We have checked that the momentum distribution is not sensitive to the truncation parameters $N_\text{max}$ and $K$. To be more specific, we observe that the calculation using $N_\text{max}=32$ and $K=32$ provides similar results comparing to calculation using $N_\text{max}=24$ and $K=24$ up to $x^\LCp=50$~GeV$^{-1}$. The numerical scheme adopted in the tBLFQ simulation is the MSD6 scheme, see Appendix \ref{apd:MSD} for details.

We show the transverse momentum distribution of the electron at $x^\LCp=20$, $30$, $40$ and $50$~GeV$^{-1}$ from top to bottom in Fig.~\ref{fig:nonpert2}. The solid black curve is the initial transverse momentum distribution, which peaks around $0.7b$ with the peak value approximately equal to $1.7$. The tBLFQ simulation predicts that the transverse momentum distribution follows the LFPT predictions at first ($x^\LCp \lesssim 20$~GeV$^{-1}$). After being exposed to the intense field for a longer time, e.g., $x^\LCp \gtrsim 30$~GeV$^{-1}$, the peak value increases according to the tBLFQ simulation while both the LO and NLO LFPT predict that the value of the peak should decrease. Moreover, the tBLFQ simulation predicts that the position of the peak would be at a lower momentum comparing to the LFPT calculations, but with a smaller width. The different predictions by the tBLFQ simulation and the LFPT calculation could potentially be used as a quantitative observable for the higher order effects in this process. In Fig.~\ref{fig:nonpert4}, we show a snapshot of the transverse momentum distribution for $b < p_\perp < 4 b$ on a semi-log scale for the same process in Fig.~\ref{fig:nonpert2} at $x^\LCp=50$~GeV$^{-1}$. It shows that the probability to find the electron with larger transverse momentum is significantly higher comparing to the initial distribution, if the electron has been exposed to the intense field for a sufficient amount of time. It is apparent because the electron has been excited to higher radial states from the initial $n=0$ state.

We show the longitudinal momentum distribution of the electron at $x^\LCp=20$, $30$, $40$ and $50$~GeV$^{-1}$ from top to bottom in Fig. \ref{fig:nonpert3}. Initially, the longitudinal momentum of the electron is $p^\LCp=17\pi/L$. Since the strong e.m. field generated by the heavy ion has a non-trivial longitudinal momentum distribution, the evolving amplitude for the electron receives contributions showing that it is both accelerated as well as decelerated in the longitudinal direction. We discuss results of the tBLFQ simulation first. After being exposed to the strong field for $20$~GeV$^{-1}$, the probability to find the electron in the initial longitudinal configuration has decreased to about $88\%$, and it is most likely to be found in states with longitudinal momentum adjacent to the initial momentum, \eg, the probabilities to find the electron with $p^\LCp=19\pi/L$ and $p^\LCp=15\pi/L$ are about $7\%$ and $1.5\%$, respectively. The transition rate to the $p^\LCp=19\pi/L$ state is larger, because the kinetic energy difference between the $p^\LCp=17\pi/L$ state and the $p^\LCp=19\pi/L$ state is smaller comparing to the kinetic energy difference between the $p^\LCp=17\pi/L$ state and the $p^\LCp=15\pi/L$ state for the same transverse momentum distribution, owing to the fact that the kinetic energy of the electron is inversely proportional to the longitudinal momentum. The probability to find the electron in the initial longitudinal configuration continues decreasing with increasing exposure time in the intense field. At $x^\LCp=50$~GeV$^{-1}$, the probability to find the electron in the initial longitudinal configuration has decreased to about $75\%$. The probabilities to find the electron in states with $p^\LCp=13\pi/L$ and $p^\LCp=21\pi/L$ are about $1\%$ and $2\%$, respectively, which are not negligible. The probabilities to find the electron with $p^\LCp=19\pi/L$ and $p^\LCp=15\pi/L$ have increased to $14\%$ and $4\%$, respectively. The probabilities to find the electron in other longitudinal momentum states are also building up over time.    

Comparing to the tBLFQ simulation, the LO LFPT calculation underestimates the depletion of the initial longitudinal momentum state. At $x^\LCp=50$~GeV$^{-1}$, the probabilities to find the electron in the initial longitudinal configuration, states with $p^\LCp=19\pi/L$ and $p^\LCp=15\pi/L$  are about $83\%$, $6\%$ and $6\%$, respectively. On the other hand, the NLO LFPT calculation predicts slightly higher transition probability to other longitudinal momentum states. At $x^\LCp=50$~GeV$^{-1}$, the probabilities to find the electron in the initial longitudinal state, and the states with $p^\LCp=19\pi/L$ and $p^\LCp=15\pi/L$  are about $71\%$, $16\%$ and $4\%$, respectively.

\section{Conclusions and Outlook}
\label{sec:conclusion}

In this paper we use the recently formulated time-dependent basis light-front quantization formalism to study the influence of an electromagnetic field generated by an ultra-relativistic nucleus on a charged fermion field. We show that the information of the system is accessible at any intermediate time at the amplitude level using the tBLFQ approach. We show the BLFQ basis calculation is compatible with the momentum basis calculation with sufficiently large truncation parameters. Further, we calculate the transition of an electron influenced by the field generated by an ultra-relativistic nucleus and it shows agreement with light-front perturbation theory when the atomic number of the ultra-relativistic nucleus is small. We find that higher-order contributions are significant for nuclei with a large atomic number. We then demonstrate that the real-time evolution of the momentum distribution of an electron evolving inside the strong electromagnetic field can be calculated non-perturbatively using the tBLFQ approach.

Next, we plan to apply tBLFQ to QCD processes in high energy nuclear collisions, following the same procedures we presented in this investigation. For example, we can take the semi-analytic solution of the quasi-classical early time gluon field created in high energy nuclear collisions as background fields \cite{Fries:2006pv, Chen:2015wia}, and study the evolution of quarks and gluons in this field. Thus we could calculate high energy jet and heavy quark modification by the early time gluon field which could lead to improved understanding of jet quenching and heavy quark physics in heavy ion collisions \cite{Qin:2015srf,Rapp:2008qc}. We also plan to apply the tBLFQ formalism to electron-ion collisions \cite{Accardi:2012qut}, to study the evolution of quarks and gluons in classical color fields. For example, we could study diffractive processes in electron-ion collisions using the dipole picture along with the classical description of a high energy nucleus from CGC, such that the dipole cross section and vector meson light-front wavefunction are obtained in a unified formalism \cite{Chen:2016dlk}. The advantages of the tBLFQ framework are distinctive: it is a non-perturbative, first-principles numerical scheme; the calculation is at the amplitude level thereby incorporating quantum interference effects; and we can naturally extend our calculation to higher Fock-sectors as well as go beyond the Eikonal approximation.

Further improvement of tBLFQ relies on the development of BLFQ itself. For example, progress on implementing a sector-dependent renormalization scheme within the BLFQ framework \cite{Zhao:2014hpa} will allow inclusion of higher Fock-sectors in our calculation; and what is more important, a proper renormalization scheme enables us to study various processes involving particle production and annihilation. For instance, we would be enabled to apply tBLFQ to the lepton pair and quarkonium production processes in ultra-peripheral heavy ion collision \cite{Baltz:2007kq,Afanasiev:2009hy,Abbas:2013oua}. As computing technology advances, we envision that tBLFQ formalism will become a tool with increasing utility.

\acknowledgments

We thank P.~Maris for valuable remarks and constructive criticisms. We acknowledge A.~Ilderton for fruitful discussions and M.~Li for checking some equations in the paper. This work was supported by the Department of Energy under Grant Nos. DE-FG02-87ER40371 and DESC0008485 (SciDAC-3/NUCLEI). X. Zhao is supported by the new faculty startup funding by the Institute of Modern Physics, Chinese Academy of Sciences. We acknowledge computational resources provided by the National Energy Research Scientific Computing Center (NERSC), which is supported by the Office of Science of the U.S. Department of Energy under Contract No. DE-AC02-05CH11231.

\appendix
\section{Conventions}
\label{apd:convention}
The conventions we use in this paper are summarized in this section. Light-front coordinates are related to covariant coordinates $(x^0,x^1,x^2,x^3)$ as follows,
\begin{equation}
x^\LCpm =x^0 \pm x^3 \, , \quad \b x^\perp=(x^1,x^2)
\end{equation}
with $x^\LCp$ regarded as light-front time, and $x^\LCm$ is the longitudinal coordinate. $x^\perp=(x^1,x^2)$ are the transverse coordinates. Non-vanishing elements of the metric tensor are $g^{+-}=g^{-+}=2$ and $g^{11}=g^{22}=-1$.

The basis states in the transverse direction are the eigenstates of the following two-dimensional harmonic oscillator (2D-HO) Hamiltonian
\begin{align}
    \label{HO_hami}
    H^{2d}_{HO}=\frac{p^2_\perp}{2M}+\frac{1}{2}M\Omega^2 x^2_\perp,
\end{align}
in which $M$ and $\Omega$ are the mass and frequency of the oscillator. The characteristic scale of the 2D-HO depends only on a combination of these two parameters which we denote as $b=\sqrt{M\Omega}$. The eigenstates of Eq.~(\ref{HO_hami}) have two quantum numbers, the radial excitation $n$, and angular momentum projection $m$. The eigenenergy  of a state with quantum number $n$ and $m$ is $E_{n,m}=(2n+|m|+1)\Omega$.

The basis wavefunctions in polar coordinates $(\rho,\phi)$, with  $x^1=\rho \cos\phi$ and $x^2=\rho \sin\phi$, are,
\begin{align}
    \label{transverse_HO_wavefunction}
    \Phi^b_{nm}(\rho,\phi)=(-1)^{n}i^{|m|}f_{nm}(\rho)\chi_m(\phi) \;,
\end{align}
where the radial part $f_{nm}(\rho)$ is expressed by generalized Laguerre polynomials, $L^{|m|}_n(b^2\rho^2)$ as,
\begin{align}
    f^b_{nm}(\rho)= b\sqrt{2}\sqrt{\frac{n!}{(n+|m|)!}}\ e^{-b^2\rho^2/2} (b\rho)^{|m|}L^{|m|}_n(b^2\rho^2)\; ,
    \end{align}
and angular part is
\begin{align}
     \chi_m(\phi)=\frac{1}{\sqrt{2\pi}}e^{im\phi} \; .
\end{align}
A Fourier transform of the HO coordinate space wavefunctions immediately gives HO wavefunctions in momentum space,
\begin{align}
    \label{HO_wavefunction_fourier_transform}
    \tilde\Phi^b_{nm}(p^\perp) =(2\pi)\tilde f^b_{nm}(p)\tilde\chi_m(\phi)\; ,
\end{align}
with
\begin{align}
    \tilde f^b_{nm}(p)=&\frac{\sqrt{2}}{b}\sqrt{\frac{n!}{(n+|m|)!}}e^{-p^2/(2b^2)} \left(\frac{p}{b}\right)^{|m|}L^{|m|}_n\left(\frac{p^2}{b^2}\right)\ ,
\end{align}
and
\begin{align}
    \tilde\chi_m(\phi)=\frac{1}{\sqrt{2\pi}}e^{im\phi}\ .
\end{align}
The coordinate and momentum space wavefunctions Eqs.~(\ref{transverse_HO_wavefunction}) and (\ref{HO_wavefunction_fourier_transform}) differ only in an overall coefficient if expressed as dimensionless parameter $b\rho$ and $p/b$.

The mode expansion of the fermion field operators in the BLFQ basis is,
\begin{align}
    \label{field_op_momentum_basis_BLFQ}
    \Psi(x) &= \sum_{\bar\alpha}\frac{1}{\sqrt{2L}} \int\! \frac{\ud^2 \bm p^\perp}{(2\pi)^2} \big[b_{\bar\alpha}\tilde\Phi_{nm}(\bm p^\perp)u(p,\lambda) e^{-i{\sf p\cdot x}} \nonumber \\
    &\qquad +d^\dagger_{\bar\alpha} \tilde\Phi^*_{nm}(\bm p^\perp)v(p,\lambda) e^{i{\sf p\cdot x}} \big]  \;,
\end{align}
where ${\sf p\cdot x}=\frac{1}{2}p^+x^\LCm-\bm p^\perp\cdot \bm x^\perp$ is the 3-product for the spatial components of $p^\mu$ and $x^\mu$.  The creation operators $b^\dagger_{\bar\alpha}$ and $d^\dagger_{\bar\alpha}$ create electrons and positrons respectively with quantum numbers $\bar\alpha=\{k,n,m,\lambda\}$. They obey the anti-commutation relations
\be
	 \{ b_{\bar\alpha},b^\dagger_{{\bar\alpha}'} \} = \{ d_{\bar\alpha},d^\dagger_{{\bar\alpha}'} \} = \delta_{\bar{\alpha}\bar{\alpha}'} \,.
\ee
We use the following (chiral) spinor representation, with helicity $\lambda=\pm 1/2$,
\begin{align}
    &u(p,\frac{1}{2})=\left(\begin{matrix}1\cr 0\cr \frac{im_e}{p^+}\cr
\frac{(ip^1-p^2)}{p^+}\cr\end{matrix}\right), \quad
    u(p,-\frac{1}{2})=\left(\begin{matrix}0\cr 1\cr \frac{(-ip^1-p^2)}{p^+}\cr \frac{im_e}{p^+}\cr
\end{matrix}\right), \quad\nonumber\\
    &v(p,\frac{1}{2})=\left(\begin{matrix}0\cr 1\cr\frac{(-ip^1-p^2)}{p^+}\cr\frac{-im_e}{p^+}\cr\end{matrix}\right), \quad
    v(p,-\frac{1}{2})=\left(\begin{matrix}1\cr 0\cr  \frac{-im_e}{p^+}\cr  \frac{(ip^1-p^2)}{p^+}\cr
\end{matrix}\right).
\label{eq:spinors}
\end{align}

\section{Multistep  Differencing Scheme}
\label{apd:MSD}
Various schemes have been proposed for solving Eq.~(\ref{i_evolve_sch}) numerically. For example, the Crank-Nicholson scheme (CN), which is unconditionally stable and accurate up to $(P^\LCm \delta x^\LCp)^2$. However it is an implicit scheme which requires matrix inversion, a non desirable feature demanding tremendous computation efforts. There is also the Chebyshev scheme which approximates the exponential function by a Chebyshev polynomial expansion, which is stable and accurate however the intermediate wavefunctions are not available.

The multistep differencing scheme is an extension of the Euler scheme, which is stable and accurate while providing intermediate wavefunctions. The second order differencing scheme (MSD2) \cite{Askar:1978} relates the state at $x^+{+}\delta x^+$ to those at $x^+$ and $x^+{-}\delta x^+$ via
\begin{align}
    \label{evolve_MSD2_scheme}
     \ket{\psi;x^+{+}\delta x^+} = & \ket{\psi;x^+{-}\delta x^+} -i P^\LCm (x^+)\delta x^+\ket{\psi;x^+} \nonumber \\
     & + \mathcal{O}\big( (P^\LCm \delta x^\LCp)^3 \big) \;.
\end{align}
It is conditionally stable if $|P^\LCm_{\text{max}}|\delta x^+<1$, where $P^\LCm_{\text{max}}$ is the largest (by magnitude) eigenvalue of $P^\LCm$ when $P^\LCm$ is time-independent~\cite{Iitaka:1994}.

It has been shown that higher order multistep differencing scheme can provide much higher accuracy with some increase of computation efforts~\cite{Iitaka:1994}.
The fourth order scheme MSD4,
\begin{align}
\ket{\psi;x^+{+} 2\delta x^+}
    \approx & \ket{\psi;x^+{-}2\delta x^+}  -2 i P^\LCm (x^+) \delta x^+   \Big [ -\frac{1}{3} \ket{\psi;x^+}
    \nonumber \\
    & + \frac{2}{3} ( \ket{\psi;x^+ {+} \delta x^+}
    +\ket{\psi;x^+ {-} \delta x^+} ) \Big ] \nonumber  \\
    &+ \mathcal{O} \big ((P^\LCm (x^+) \delta x^+)^5 \big)  \; ,
\end{align}
which is accurate up to $(P^\LCm \delta x^+)^4$ and stable if $|P^\LCm_{\text{max}}|\delta x^+<0.4$. The sixth order scheme MSD6,
\begin{align}
\ket{\psi;x^+{+} 3\delta x^+}
    \approx &\ket{\psi;x^+{-}3\delta x^+}  - 3 i P^\LCm (x^+) \delta x^+   \Big [ \frac{13}{10} \ket{\psi;x^+}
    \nonumber \\
    & - \frac{7}{10} ( \ket{\psi;x^+ {+} \delta x^+}
    +\ket{\psi;x^+ {-} \delta x^+} )
    \nonumber \\
    &   +  \frac{11}{20} ( \ket{\psi;x^+ {+} 2\delta x^+}
    +\ket{\psi;x^+ {-} 2\delta x^+} )      \Big ] \nonumber \\
    &+ \mathcal{O} \big ((P^\LCm (x^+) \delta x^+)^7 \big)
\end{align}
it is accurate up to $(P^\LCm \delta x^+)^6$ and stable when $|P^\LCm_{\text{max}}|\delta x^+<0.1$.

The accuracy of MSD6 scheme for the calculation performed in Section \ref{sec:result} can be checked by comparing the evolution of eigenstates $\ket{B}$ of the Hamiltonian in Eq.~(\ref{eq:H_TandV}), \ie, $P^\LCm \ket{B} = P^\LCm_B \ket{B}$, which is just a phase factor $\exp(i P^\LCm_B \Delta x^\LCp)$. Note we use $\ket{B}$ to avoid confusion with eigenstate $\ket{\beta}$ of $P^\LCm_\text{QED}$. The calculations in the BLFQ basis are tested to be accurate up to $4$ significant figures at $x^\LCp = 50$~GeV$^{-1}$ by successively halving the time increment. 


%
\section{QED Hamiltonian in the BLFQ basis}
\label{apd:Ham}

\begin{table}[ht!]
    \renewcommand{\arraystretch}{2.5}
    \centering
\begin{tabular}{|c|c|}
\hline
($\lambda_2$,$\lambda_1$) & $\bar{u}(p_2,\lambda_2)\gamma^\mu u(p_1,\lambda_1) \mathcal{A}_\mu(k)$ \\ \hline
$\uparrow \uparrow$ & $\frac{Z\vert e \vert}{\epsilon\left(  e^{-2y} (k^\LCp)^2 +  k_\perp^2 \right)} \big( 2e^{-2y} + \frac{\bar p_1 \bar k^*}{ p^\LCp_1  k^\LCp}+\frac{\bar p_1^* \bar k}{p^\LCp_2 k^\LCp} \big)$\\ \hline
$\uparrow \downarrow$ & $\frac{Z\vert e \vert}{\epsilon \left(  e^{-2y} (k^\LCp)^2 +  k_\perp^2 \right) } \frac{m_e \bar k^*}{ k^\LCp}\big( \frac{1}{p_1^\LCp}-  \frac{1}{p_2^\LCp} \big)$   \\ \hline
$\downarrow \uparrow$ & $\frac{Z \vert e \vert}{\epsilon \left(  e^{-2y} (k^\LCp)^2 +  k_\perp^2 \right) } \frac{m_e \bar k}{ k^\LCp}\big( \frac{1}{p_2^\LCp}-  \frac{1}{p_1^\LCp} \big)$ \\ \hline
$\downarrow\downarrow$ & $\frac{Z \vert e \vert}{\epsilon\left(  e^{-2y} (k^\LCp)^2 +  k_\perp^2 \right)} \big( 2e^{-2y} + \frac{\bar p_1^* \bar k}{ p^\LCp_1  k^\LCp}+\frac{\bar p_1 \bar k^*}{p^\LCp_2 k^\LCp} \big)$ \\ \hline
($\lambda_2$,$\lambda_1$) & $\bar{u}(p_2,\lambda_2)\gamma^i \gamma^\LCp \gamma^j u(p_1,\lambda_1) \mathcal{A}_i (k_2) \mathcal{A}_j (k_1)$ \\ \hline
$\uparrow\uparrow$ & $ (\frac{Z \vert e \vert}{\epsilon})^2 \frac{2 \bar k_1^* \bar k_2  }{\left(  e^{-2y} (k_1^\LCp)^2 +  {k_1^{\perp}}^2 \right)\left(  e^{-2y} (k_2^\LCp)^2 +  {k_2^{\perp}}^2 \right) }$\\ \hline
$\uparrow\downarrow$ & 0 \\ \hline
$\downarrow\uparrow$ & 0\\ \hline
$\downarrow\downarrow$ & $ (\frac{Z \vert e \vert}{\epsilon})^2 \frac{2 \bar k_1^* \bar k_2  }{\left(  e^{-2y} (k_1^\LCp)^2 +  {k_1^{\perp}}^2 \right)\left(  e^{-2y} (k_2^\LCp)^2 +  {k_2^{\perp}}^2 \right) }$ \\ \hline
\end{tabular}
\caption{Spinor background field potential vector contraction for different helicity configurations of the incoming electron (``1'') and the outgoing electron (``2''). We define the complex momentum as $\bar p = p^x + i p^y $.}
\label{tab:spinor_JA_contraction_list}
\end{table}

The Hamiltonian relevant to the calculation we perform is summarized in Eq.~(\ref{eq:H_TandV}). They are fermion kinetic energy $T_f$, {\it vertex interaction} between (anti-) fermion and background fields $\mathcal{W}_1$, and {\it instantaneous-fermion interaction} between (anti-)fermion and background fields $\mathcal{W}_2$. $T_f$ in BLFQ basis has been discussed in \cite{Zhao:2013cma}. Here we outline how to express $\mathcal{W}_1$ and $\mathcal{W}_2$ algebraically in BLFQ basis following the convention in Appendix \ref{apd:convention}.

The {\it vertex interaction} between fermion and background fields $\mathcal{A}^\mu$ as in Eqs.~(\ref{eq:4vector}) is,
\begin{align}
\mathcal{W}_1 = & e\int^L_{-L}\!\ud x^-\!\! \int\!\ud^2x^\perp \bar\Psi\gamma^\mu\Psi \mathcal{A}_\mu \nonumber \\
=&\frac{e}{(2\pi)^4 L}\sum_{\bar\alpha_1,\bar\alpha_2,k^\LCp} \int\! \ud^2(\bm p^\perp_{1},\bm p^\perp_{2},\bm k^\perp) \nonumber \\
     &\times  \tilde\Phi^*_{n_2 m_2}(\bm p^\perp_2) \tilde\Phi_{n_1 m_1}(\bm p^\perp_1) \nonumber \\
     &\times \bar{u}(p_2,\lambda_2)\gamma^\mu u(p_1,\lambda_1) \mathcal{A}_\mu(x^\LCp,k)
     \nonumber \\
     &\times \delta^{(2)}(\bm p^\perp_2-\bm p^\perp_1-\bm k^\perp)  \delta(p_2^\LCp \vert p_1^\LCp+k^\LCp)b^\dagger_{\bar\alpha_2} b_{\bar\alpha_1}  \; ,
     \label{eq:W1}
\end{align}
where $\bar\alpha_1,\bar\alpha_2$ are the quantum numbers associated with the field operators  $\Psi$ and $\bar\Psi$ respectively. $k=(k^\LCp,\bm k^\perp)$ is the momentum 3-vector of the background fields. We have in the transverse direction the 2D Dirac delta function and the Kronecker delta for the discretized longitudinal momentum. The {\it instantaneous-fermion interaction} between fermion and background fields $\mathcal{A}^\mu$ as in Eqs.~(\ref{eq:4vector}) is,
\begin{align}
\mathcal{W}_2 = &  \frac{e^2}{2} \int_{-L}^{L} \ud x^\LCm \int \ud^2 x^\perp   \bar{\Psi} \gamma^i \mathcal{A}_i \frac{\gamma^\LCp}{i\partial^\LCp} \gamma^j \mathcal{A}_j \Psi  \,  \nonumber \\
=&\frac{e^2}{2 (2\pi)^8 L^2} \sum_{\bar\alpha_1,\bar\alpha_2,k_1^\LCp,k_2^\LCp,n_i,m_i} \int\! \ud^2(\bm p^\perp_{1},\bm p^\perp_{2},\bm k_1^\perp,\bm k_2^\perp) \nonumber \\
&\times      \tilde\Phi^*_{n_2 m_2}(\bm p^\perp_2) \tilde\Phi_{n_i m_i}(\bm p_2^\perp - \bm k_2^\perp) \nonumber \\
&\times \tilde\Phi^*_{n_i m_i}(\bm p^\perp_1+\bm k_1^\perp) \tilde\Phi_{n_1 m_1}(\bm p^\perp_1)
\nonumber \\
&\times \frac{ \bar{u}(p_2,\lambda_2) \gamma^i \gamma^\LCp \gamma^j u(p_1,\lambda_1)}{p_1^\LCp + k_1^\LCp} \mathcal{A}_i (x^\LCp,k_2) \mathcal{A}_j (x^\LCp,k_1) \nonumber \\
&  \times  \delta(p_2^\LCp \vert p_1^\LCp+k_1^\LCp+k_2^\LCp) b^\dagger_{\bar\alpha_2} b_{\bar\alpha_1}  \; .
     \label{eq:W2}
\end{align}
and we have used the HO wavefunction representation of the Dirac delta function in the transverse direction,
\begin{equation}
(2\pi)^2 \delta^{(2)} (\bm p_2^\perp - \bm p_1^\perp )  = \sum_{n_i,m_i} \tilde\Phi_{n_i,m_i}(\bm p_2^\perp)  \tilde\Phi^*_{n_i,m_i}(p_1^\perp)   \; ,
\end{equation}
so the integration with respect to $\bm p^\perp_{1},\bm p^\perp_{2},\bm k_1^\perp,\bm k_2^\perp$ can be factorized. The delta function is exact only if $n_i$ and $m_i$ are summed over all possible values. However in LFPT, $\mathcal{W}_2$ has a singularity and it is canceled by the second order vertex interaction. Which implies we should regard $n_i$ and $m_i$ as quantum numbers of an intermediate fermion line and should be subjected to the same $N_\text{max}$ truncation constraint.

We list the spinor background potential contraction for different fermion helicities in Table \ref{tab:spinor_JA_contraction_list}. Note that the exponential phase factors in Eq.~(\ref{eq:4vector}) are suppressed in the table.

Integration over the product of more than one highly oscillatory, 2D-HO wavefunctions, as in Eq.~(\ref{eq:W1},\ref{eq:W2}) can be simplified by applying the Talmi-Moshinsky transformation to the 2D-HO wavefunctions~\cite{Davies}. Eventually we are dealing with integration,
\begin{equation}
\int\! \ud^2 p^\perp \tilde \Phi_{n m}(p^\perp) \frac{1}{{\bm p^\perp}^2+ e^{-2y} (p^\LCp)^2}
\end{equation}
which can be calculated as a finite-term summation using a series expansion of the Laguerre polynomials.

\end{document}